\documentclass[aps,prb,twocolumn,showpacs]{revtex4}
\usepackage{txfonts}
\usepackage{hyperref}
\usepackage{graphicx}
\usepackage{color}
\hypersetup{ colorlinks,
linkcolor=blue,
filecolor=blue,
urlcolor=blue,
citecolor=red}

\begin{document}
\title{A novel high resolution contactless technique for thermal field mapping and thermal conductivity determination:  Two-Laser Raman Thermometry}
\author{J. S. Reparaz$\,^{1}$}
\author{E. Chavez-Angel$\,^{1,2}$}
\author{M. R. Wagner$\,^{1}$}
\author{B. Graczykowski$\,^1$}
\author{J. Gomis-Bresco$\,^{1}$}
\author{F. Alzina$\,^1$}
\author{C. M. Sotomayor Torres$\,^{1,3}$}

\address{$\,^1$ICN2 - Institut Catala de Nanociencia i Nanotecnologia, Campus UAB, 08193 Bellaterra (Barcelona), Spain}
\address{$\,^2$Dept. of Physics, UAB, 08193 Bellaterra (Barcelona), Spain}
\address{$\,^3$ICREA, Passeig Llu\'is Companys 23, 08010 Barcelona, Spain}

\begin{abstract}
\noindent We present a novel high resolution contactless technique for thermal conductivity determination and thermal field mapping 
based on creating a thermal distribution of phonons using a heating laser, while a second laser probes the local temperature through
the spectral position of a Raman active mode.
The spatial resolution can be as small as $300$ nm, whereas its temperature accuracy is $\pm 2$ K.
We validate this technique investigating the thermal properties of three free-standing single crystalline Si membranes with thickness of 250, 1000, and 2000 nm. 
We show that for 2-dimensional materials such as free-standing membranes or thin films, and for small temperature gradients,  the thermal 
field decays as $T(r) \propto ln(r)$ in the diffusive limit.
The case of large temperature gradients within the membranes leads to an exponential decay of the thermal field, 
$T \propto exp[-A \cdot ln(r)]$.
The results demonstrate the full potential of this new contactless method for quantitative determination of thermal properties. 
The range of materials to which this method is applicable reaches far beyond the here demonstrated case of Si, as the only
requirement is the presence of a Raman active mode.


\end{abstract}
\maketitle
\section{Introduction}

A precise determination of the thermal conductivity ($\kappa$) of a given material is usually a difficult task since the heat per unit time flowing in a certain 
spatial direction ($Q \equiv P_{abs}$) must be precisely determined. Contrary to the analogous case of electrons (or holes) propagating in a metal, where the electrical current
can be easily measured using galvanometer through the magnetic field created by the current, there is no direct method to measure heat currents.
Instead, the heat flux per unit time is usually inferred considering the geometry of the system and the excitation source, 
the temperature gradient ($\Delta T$) is measured in the heat flow direction and, finally,
the thermal conductivity is obtained through Fourier's law ($\kappa=P_{abs}/\Delta T$). 
Several electrical and optical techniques have been developed to measure the thermal conductivity of a 
large variety of materials and structures.\cite{Parker1961,Hatta1985, Paddock1986,Schmotz2010,Saltonstall2013,Tsu1982,Williams1986,Schmidt2009,Cahill1990, Harata1990,Capinski1996,Govorkov1997,Johnson2012}
Optical methods have recently attracted considerable attention 
since most of them are contactless and, thus, require few sample preparation.
Such techniques can be divided into two main categories: i) steady-state techniques,\cite{Schmotz2010, Saltonstall2013, Tsu1982,Williams1986,Govorkov1997,Schmidt2009} and 
ii) transient techniques.\cite{Parker1961,Paddock1986,Harata1990,Capinski1996, Johnson2012} 
An advantage of transient techniques is that they do not require the knowledge of the absorbed power in the sample since they are only sensitive to the thermal diffusivity,
$\alpha=\kappa/\rho C_p$ (where $\rho$ is the density and $C_p$ the specific heat), which is usually proportional to a characteristic decay time of the system. 
On the contrary, steady-state techniques provide a direct determination of the thermal conductivity
provided that the power absorbed in the sample is known.
Depending on the materials and structures under investigation either of these approaches might be more convenient.

Here we present a novel high resolution contactless technique suitable for thermal conductivity determination and thermal mapping of 
1- and 2-dimensional structures provided that the samples exhibit a detectable Raman signal of any of its active modes, and
a sufficiently large spectral shift as function of temperature.
The thermal conductivity is obtained by fitting the spatial decay of the thermal field using a diffusive model based on Fourier's law.
We provide the solution of the 2-dimensional steady-state heat equation for an isotropic medium for the cases where the thermal conductivity is 
constant ($\kappa_0$) or varies with  temperature ($\kappa(T)$). 
The thermal properties of three free-standing Si membranes with thicknesses of 250, 1000, and 2000 nm were studied to validate the present approach. 

\section{Description of the technique: Two-Laser Raman Thermometry}

We have developed a contactless technique suitable to measure the thermal conductivity  of nanowires, free-standing membranes, thin films on a given substrate, and bulk samples.
This technique is based on previous works\cite{Tsu1982,Perichon1999,Perichon2000,Huang2009,Liu2011, Chavez2013, Balandin2008} which used the temperature dependent Raman spectra of a given
material to obtain its thermal conductivity (commonly known as Raman thermometry).
Briefly, the temperature rise ($T_{max}$) at the focused laser spot, which depends on the incident laser power,
is obtained from the spectral position of any Raman mode, provided that a previous
calibration of the Raman shift with temperature has been made. Subsequently, the steady-state heat equation is solved analytically or numerically, 
depending on the geometry and dimensionality of the problem, to obtain
the relationship between $\kappa$, $T_{max}$, and the absorbed power on the sample ($P_{abs}$).
The main drawback of this technique is that only one point of the temperature field is probed to determine $\kappa$ since 
the heating and the probe lasers are the same.
We recall that for the case of diffusive thermal transport at least two points are required to obtain the temperature field and, thus, $\kappa$.
To fulfill this requirement an assumption must be made, i.e. at sufficiently large distances from the laser spot the temperature equals 
that of the thermal bath ($T_{bath}$).
Thus, the two points required to obtain the temperature field are: $\{T=T_{max};{\bf r}=0\}$ and $\{T=T_{bath};\bf r\rightarrow  \infty\}$.
Although in some cases this approximation is valid such as, e.g., in the case of bulk samples, in most cases it fails leading 
to a inaccurate determination of  $\kappa$. 

\begin{figure}[t]
\includegraphics[scale=0.34]{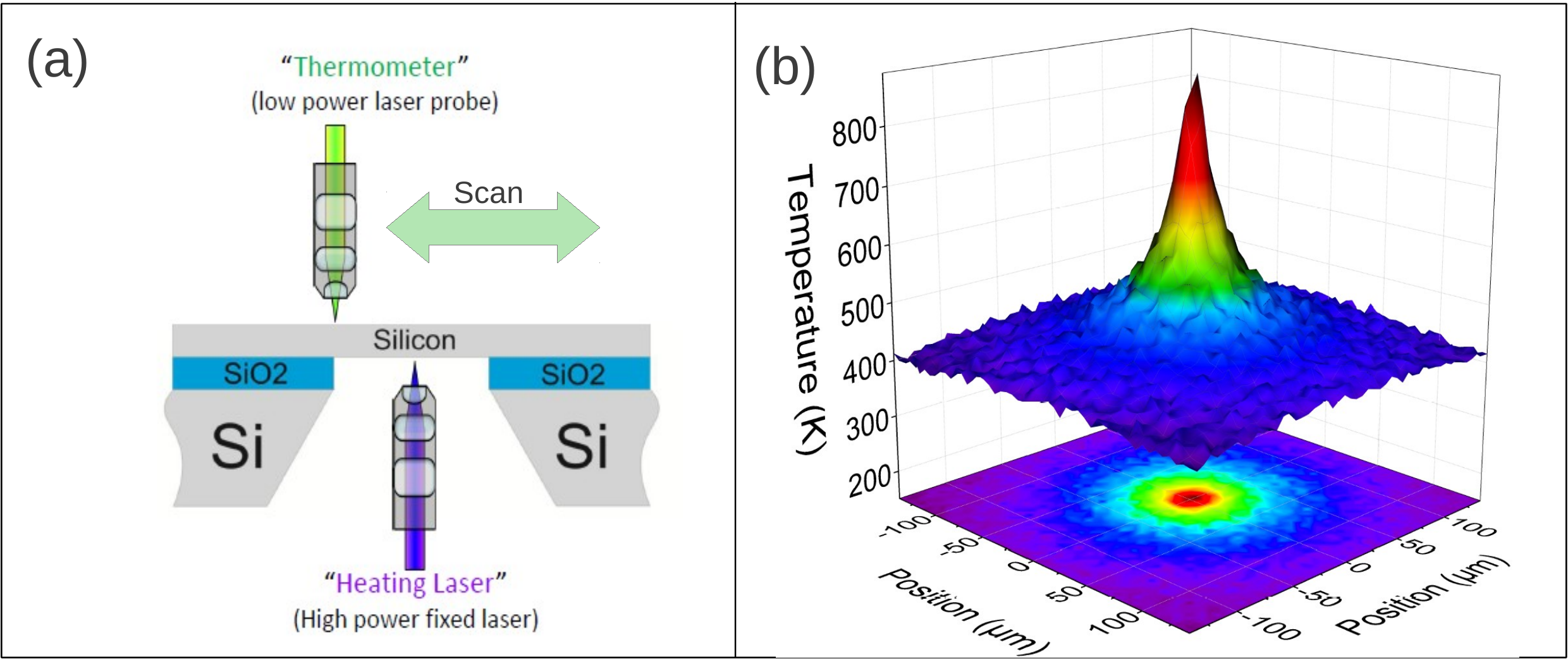} 
\caption{(a) Schematics of the Two-Laser Raman Thermometry experimental setup. To lower laser is used as heating source, whereas the upper laser probes the local temperature through the spectral shift of the longitudinal optical Raman mode of Si. (b) 2-dimensional thermal map of a 250 nm thick free-standing Si membrane. A projection of the thermal field
is also shown in a lower plane.}
\label{fig1}
\end{figure}

The Two-Laser Raman Thermometry (2LRT) technique overcomes this drawback by mapping with a low power probe laser the entire thermal field 
temperature distribution created by a high power laser impinging on the sample.
In order to apply this technique, the material under investigation ought to meet the following requirements:

\begin{itemize}
 \item {\bf Temperature Sensor.} A Raman active mode must be detectable in order to probe the local temperature. 
  This mode will ideally exhibit a strong temperature dependence ($\partial \omega /\partial T$) to be a good  temperature sensor.
  A calibration of the Raman shift versus temperature must be known in advance.
 \item {\bf Excitation Source.} The sample should have a reasonable absorbance in the spectral region of the heating laser
   to set up the temperature distribution. The absorbed power ($P_{abs}$) needs to be determined for each experimental conditions.
 \item {\bf Dimensionality.} The sample must be of 1- or 2-dimensions with lateral size larger than the optical resolution ($\approx 300$ nm). 
 It is also possible to study 3-dimensional structures provided that the heat distribution has a spherical symmetry with respect to the heating laser spot.
\end{itemize}

The main advantage of this technique as compared to other contactless steady-state methods, such as infrared thermometry, is its sub-micrometer spatial resolution.
Techniques based on scanning probes such as Scanning Thermal Microscopy (SThM), or Infrared Near Field Scanning Optical Microscopy (IR-NSOM)
also provide high spatial resolution, but their main drawback is the rather cumbersome tip calibration. On the other hand,
time-resolved techniques such as Time-Domain Thermoreflectance (TDTR) or Thermal Transient Grating (TTG) only give indirect access to 
$\kappa$ through the thermal diffusivity ($\alpha=\kappa/\rho C_P$). In particular, we mention a recent work by Schmotz {\it et al.}\cite{Schmotz2010}
where the authors measure temperature maps of a Si membrane by tracking the local change in the absorption coefficient, which exhibit
a Fabry-P\'erot interference behavior due to the sub-wavelength thickness of the membranes. Although this technique is extremely accurate 
regarding temperature determination accurate ($\delta{T} <1$ mK), it is limited to semitransparent samples, thin enough to 
produce an interference pattern. The Two-Laser Raman Thermometry technique reported here is perhaps a more general approach since it is based on the Raman effect, which is
determined by atomic vibrations within a unit cell.

In Fig. \ref{fig1}a we show schematically the 2LRT experimental arrangement. A heating laser with $\lambda_1$=405 nm is focused onto the lower surface of the Si
membranes, whereas a probe laser with $\lambda_2$=488 nm is scanned over its upper surface to obtain the local temperature.
We note that while relatively high powers were used for the heating laser ($\lambda_1$) in order to create a spatially dependent thermal field, 
low powers were employed for the probe laser ($\lambda_2$) to avoid an additional thermal perturbation. We typically used a 10:1 ratio between the heating
and probe laser powers. Both lasers are focused onto the samples using long distance 50x objectives with numerical aperture of NA=0.55. 
Although these objectives provide lower spatial resolution than short distances objectives, with typically NA$\ge$0.7, 
they allow to perform the thermal maps in a controlled environment or under variable temperature conditions.
Figure \ref{fig1}b displays a 2-dimensional (2D) temperature map of a 250 nm thick Si membrane obtained using 2LRT. The maximum temperature at the center, i.e. with the heating and probe
lasers at the same position of the membrane, is  $\approx$800 K and a radially symmetric thermal decay is observed in a constant temperature projection plane. 
The radial symmetry observed in the 2D maps arises from the isotropic thermal behavior of Si at room temperature.\cite{comment_1} However, for materials with a spatially-dependent thermal conductivity ($\kappa_{ij}$), an asymmetric thermal decay is expected. The symmetric heat distribution observed in Fig. \ref{fig1} justifies the assumption made to derive Eqs. \ref{eq1} and \ref{eq2} and reduces the measurement to only one line scan on the $(X,Y,0)$ plane containing the origin $(X,Y,Z)=(0,0,0)$. 
It is noteworthy that the temperature field does not fully
decay to the thermal bath temperature of 294 K, and instead reaches only $\approx$400 K at 150 $\mu m$ from the heating laser. The origin of this slow thermal decay is the relatively high thermal conductivity of this membrane, as well as the purely diffusive thermal transport regime. 
As shown by Eq. \ref{eq1}, or simply by Fourier's law, a larger thermal conductivity implies that heat distributes more 
efficiently through the material, thus leading to smaller temperature gradients. 
We note that a ballistic transport regime would account for a faster thermal decay as shown in Refs. [\onlinecite{Karvonen2011,ToBePub}]. 

\begin{figure}[top]
\includegraphics[scale=0.3]{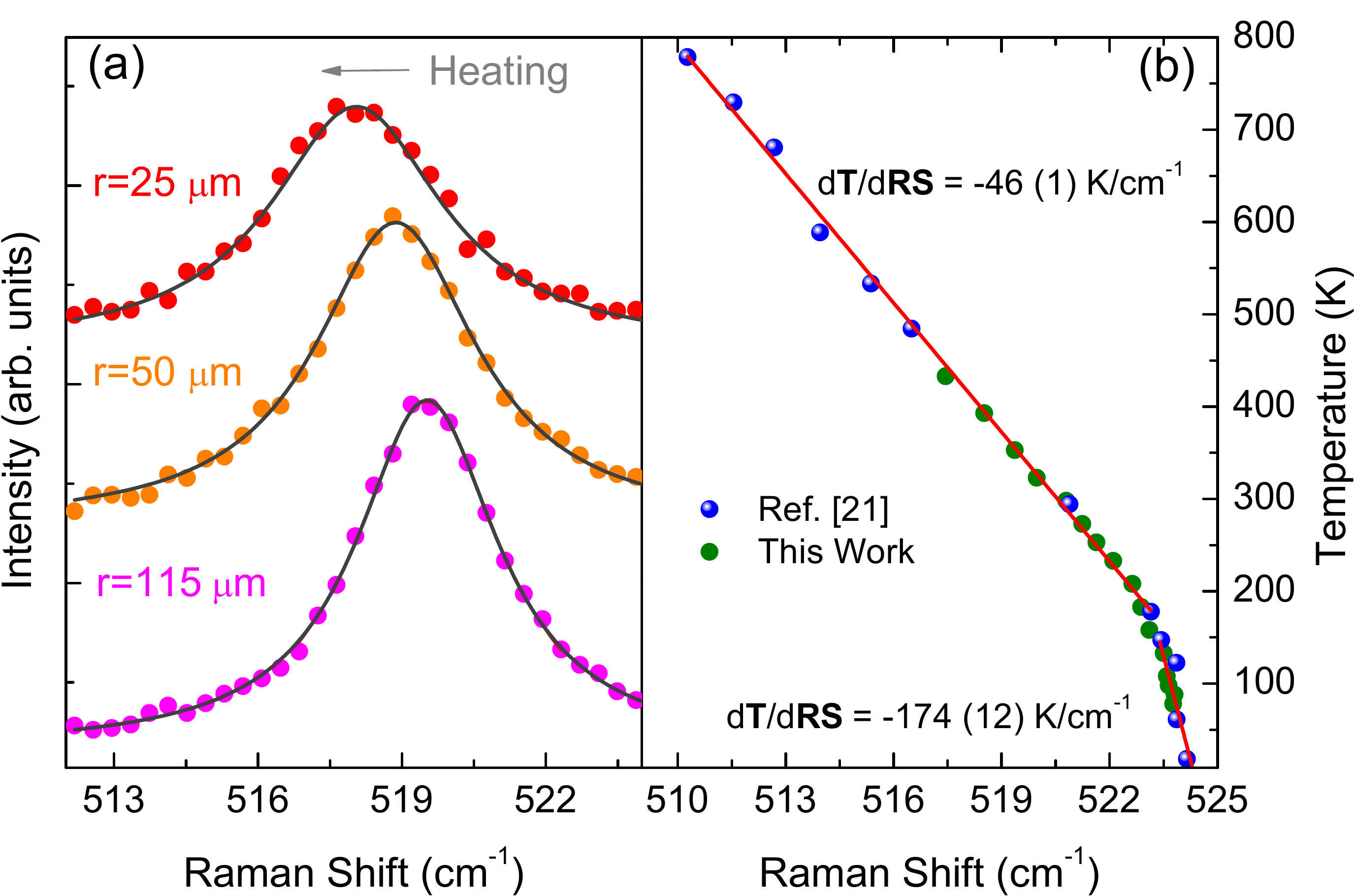} 
\caption{(a) Representative Raman spectra recorded at different distances ($r$) from the heating laser. The solid lines are least squares fits using Lorentzian functions.
(b) Calibration of the Raman shift versus temperature. The blue dots were extracted from Ref. \onlinecite{Menendez1984}}
\label{figaux}
\end{figure}

Taking into account the geometry of the measurements in Fig. \ref{fig1} we derive the analytical solution for the temperature field in the case of a free-standing isotropic membrane. 
A  temperature distribution is created upon excitation with a point like laser source as shown in Fig. \ref{fig1}.
The solution is simply given by integrating Fourier's equation: $P_{abs}/A=-\kappa \nabla T$, where $P_{abs}$ is the power 
injected in the system, $A$ is the cross sectional area of the heat flux, $\kappa$ is the thermal conductivity, 
and $T$ is the temperature. The cross sectional area depends on the geometry of the heat flux and is $A=2\pi rd$ for free-standing thin membranes
with a point-like heating source, 
where $d$ is the thickness of the membrane.
% Integrating the Fourier's equation in $r$ leads to the following solutions for the temperature field.

\begin{eqnarray}
T(r)=T_0 - \frac{P_{abs}}{2\pi d\kappa_0}\ln(r/r_0) &\rightarrow& \ \kappa=\kappa_0 \label{eq1}\\
T(r)=T_0 \left( \frac{r}{r_0} \right)^{-P_{abs}/2\pi da} &\rightarrow& \kappa=\frac{a} {T} \label{eq2}
\end{eqnarray}

\noindent where we have considered a temperature independent thermal conductivity ($\kappa_0$) as well as a $1/T$ behavior ($\kappa(T)$), 
which is typical for most semiconductors at high temperatures.\cite{Glassbrennen1964} The constant $a$ can be simply written as $a=294 \cdot \kappa(T=294)$.
We note that the spatial distribution of the heat source, i.e. a Gaussian heat source, can be considered as $P=\int_{spot} P|r-r'|dr'$, 
but we avoid this approach since it does not provide any benefits and obscures the simplicity of equations \ref{eq1} and \ref{eq2}. 
In the present case of Si use as thermometer the temperature dependent Raman shift of  the  zone center longitudinal optical (LO) phonon at $\omega_0=$ 520 cm$^{-1}$ (at 300 K).
Fig. \ref{figaux}a shows typical Raman spectra of the 250 nm thick membrane at different relative positions between the heating and probe lasers.
We recall that typically first order Raman modes in semiconductors exhibit a redshift with increasing temperatures with temperature coefficients ranging 
between 20 and 65 K/cm$^{-1}$.\cite{Menendez1984,Calizo2007,Irmer1996, Liu2000}
The Si LO peak gradually redshifts and broadens as the position of the probe laser is closer to the heating laser, i.e. smaller values of r,  indicating an increase of the local temperature.
The accuracy of the temperature determination is given by the spectral resolution. In our case, the resolving power of the spectrometer used is about 0.4 cm$^{-1}$ whereas the spectral resolution resulting from fits to the spectra using a Lorentzian function
is 0.05 cm$^{-1}$. In Fig. \ref{figaux}b we show the calibration between Raman shift and temperature as obtained using a cryostat to control the bath temperature
as well as the data from Ref. \onlinecite{Menendez1984}.
A linear relation is found between 200 and 800 K with a slope $dT/dRS=-46(1)$ K/cm$^{-1}$, which leads to a temperature accuracy
of $\pm 2$ K considering the spectral resolution. The temperature regime $T<150$ K was also fitted using a linear function obtaining
$dT/dRS=-174(12)$ K/cm$^{-1}$.

\begin{figure}[top]
\includegraphics[scale=0.4]{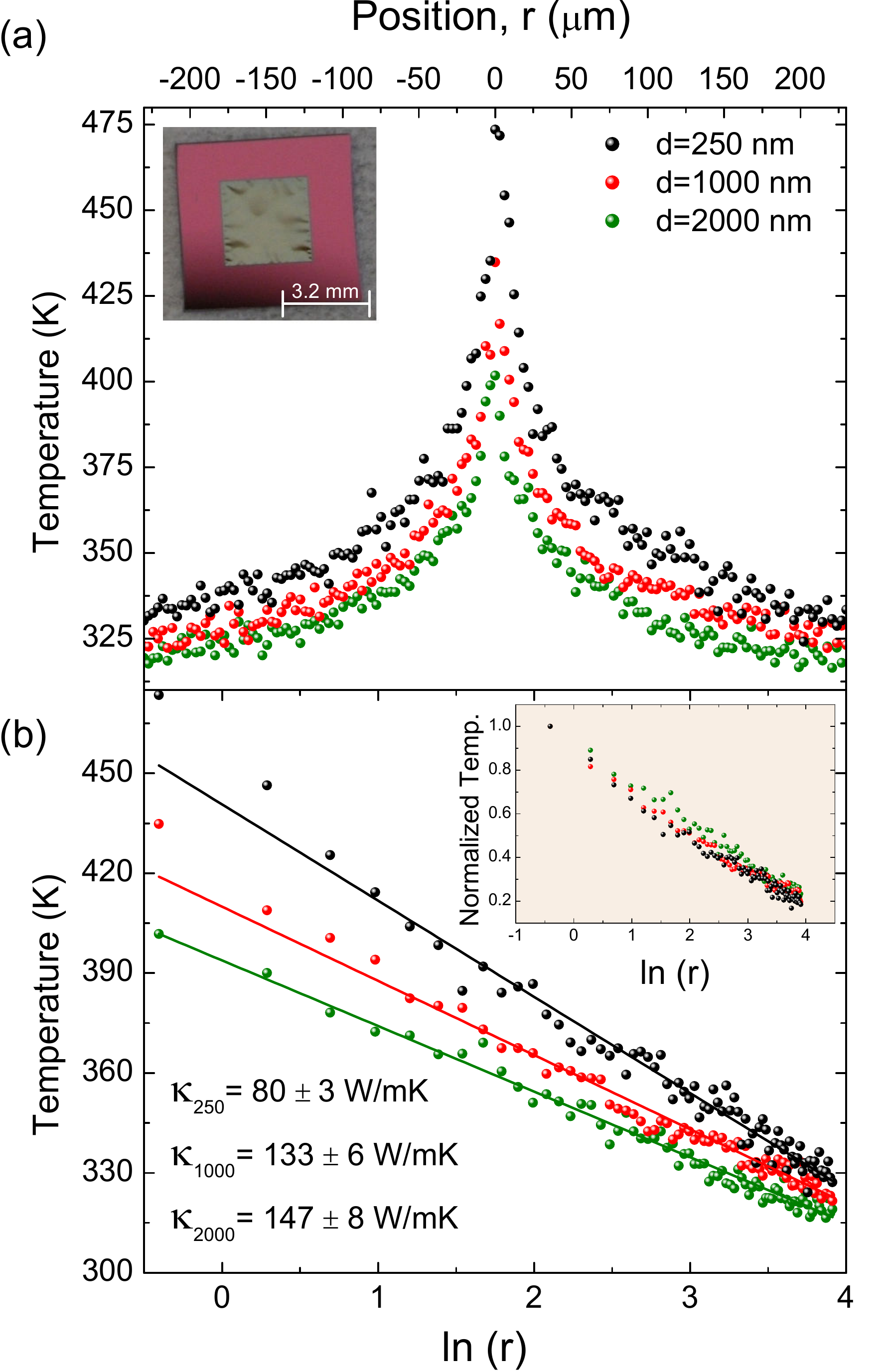} 
\caption{(a) Line scans of the thermal field as shown in Fig. \ref{fig1}b. 
The line scan was recorded using a lower heating power than for the 2D map in Fig. \ref{fig1}b to ensure that $\kappa \neq \kappa(T)$.
The inset shows an optical image of the 250 nm thick membrane. (b) Same as (a) but in logarithmic scale to visualize the ln(r) dependence
as predicted by Eq. \ref{eq1}. The inset shows the normalized temperature rise, $T_{rise}=T(r)-$294 K for the three membranes.}
\label{fig2}
\end{figure}

\section{RESULTS AND DISCUSSION}
In order to obtain the thermal conductivity of the membranes we measured a line scan of the thermal field in the $(X,Y,0)$ plane 
containing the coordinates origin as shown in Fig. \ref{fig2}a for three Si membranes with thicknesses of 250, 1000, and 2000 nm. The membranes were single crystalline 
with a surface roughness R$_a$=0.2 nm and RMS=0.15 nm, and  were purchased from NORCADA Inc. The inset shows an optical image of the 250 nm thick membrane.
All three membranes exhibit a qualitatively similar behavior, 
i.e. a similar decay length
and half width of the thermal field profile. We note that the minimum temperature obtained in these maps is also above the thermal bath temperature, reaching $\approx$330 K at 150 $\mu m$ in comparison to the $\approx$400 K obtained in the 2D thermal map of Fig. \ref{fig1}b, which arises from the smaller heating powers used for the line scans of Fig. \ref{fig2}.
This is also reflected in the maximum temperature rise observed at the central position of the line scans. 
According to Eq. \ref{eq1} the thermal decay is linear in $ln(r)$, thus, we show in Fig. \ref{fig2}b the data corresponding to Fig. \ref{fig2}a in logarithmic scale.
As predicted by Eq. \ref{eq1} the thermal field decays linearly with a slope given by  $\partial T(r)/\partial ln(r)=P_{abs}/2\pi d \kappa_0$. 
The different slopes in Fig. \ref{fig2}b simply result from different combinations of $P_{abs}$, $d$, and $\kappa_0$ with no scaling law involving $d$ or $\kappa_0$.
In fact, normalization of the temperature rise, $T_{rise}=T(r)-T_{bath}=T(r)-$294 K, results in equal slopes for the three membranes as predicted by Eq. \ref{eq1} 
(see inset of Fig. \ref{fig2}b).
Finally, using Eq. \ref{eq1} we calculate the thermal conductivity of each membrane obtained from the slopes fitted to the data in Fig. \ref{fig2}b.
We obtain $\kappa=80\pm3, 133\pm6, 147\pm8$ W/mK for the 250, 1000, and 2000 nm thick membranes, respectively, which are in excellent agreement with the results reported
in Refs. [\onlinecite{Liu2011,Asheghi1997,Asheghi1998,Ju1998, Liu2005}].

\begin{figure}[top]
\includegraphics[scale=0.45]{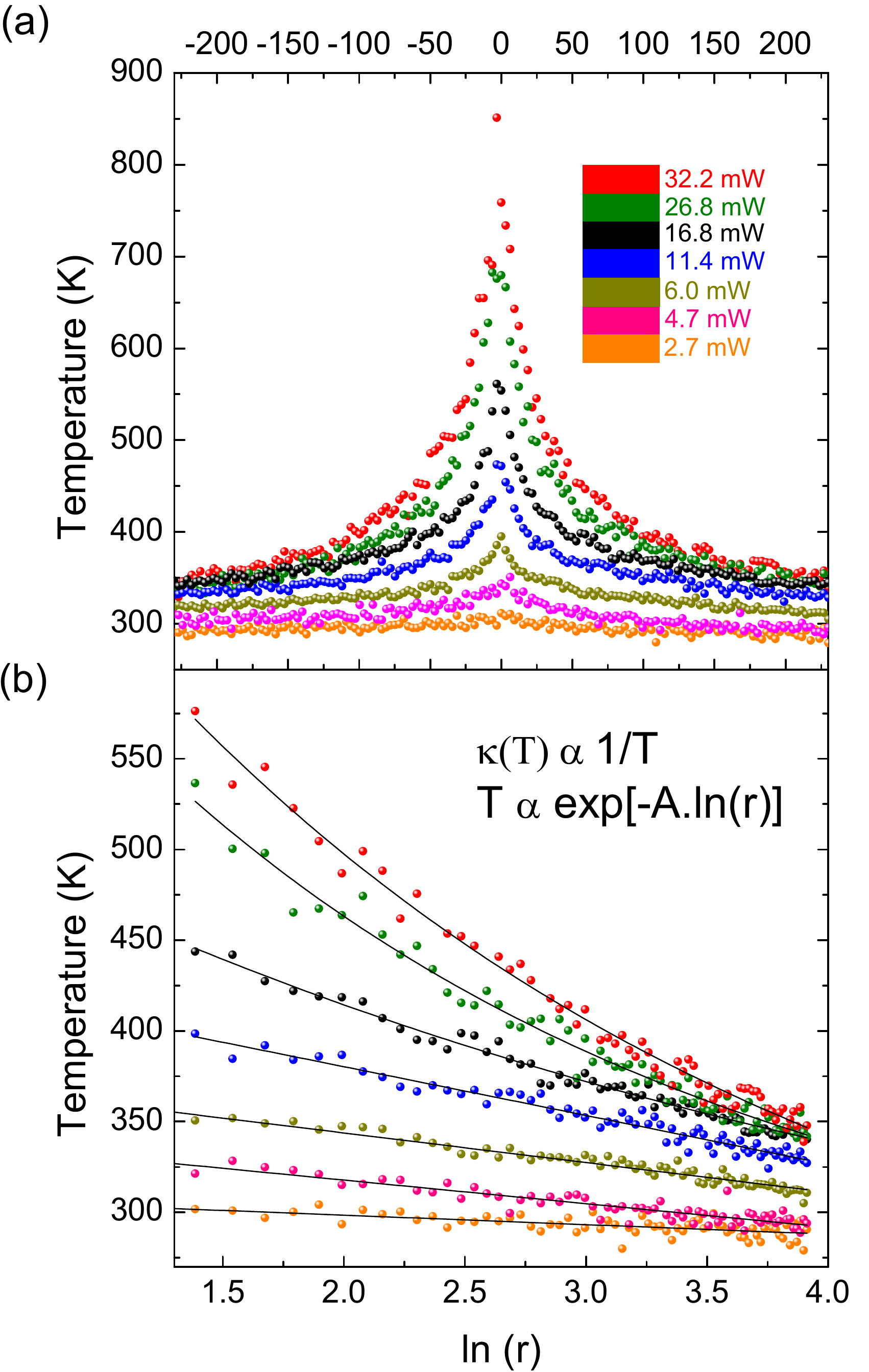} 
\caption{(a) Thermal line scans using the different heating powers for the 250 nm thick membrane. (b) Same as (a) but in logarithmic scale
to observe the deviation from the $T\propto ln(r)$ behavior given by Eq. \ref{eq1}.}
\label{fig3}
\end{figure}

The influence of a temperature dependent thermal conductivity on the line shape of the thermal field is given by Eq. \ref{eq2}. The typical temperature dependence
of the thermal conductivity for most semiconductors at high temperatures ($T > 200$ K) is $\kappa(T)=a/T^n$, where $a$ is a material dependent constant and $n$ is usually close to 1. Herein, we assume the case $n=1$ for simplicity since our main purpose is to show the effect of temperature on the decay of the 
thermal field. Figure \ref{fig3} displays several line scans using different heating powers for the 250 nm thick membrane. The maximum temperature rise 
gradually increases with increasing heating power as predicted by Eq. \ref{eq1} and \ref{eq2}. Once more, a qualitatively similar line width and decay of the thermal field profile as function the heating power is observed in Fig. \ref{fig3}a. However, a detailed analysis of the thermal decay exposes the dependence predicted by Eq. \ref{eq2} as we show in Fig. \ref{fig3}b. We recall that in
logarithmic scale and for low heating powers, i.e. $\kappa=\kappa_0 \neq \kappa(T)$, the thermal field decays as $T \propto ln(r)$ as shown by Eq. \ref{eq1}, 
whereas considering $\kappa(T)=a/T$, an exponential dependence on distance $T \propto exp[-A\cdot ln(r)]$ is expected. In fact, this is clearly observed in Fig. \ref{fig3}b
with increasing heating power.
The thermal field gradually deviates from a straight line in logarithmic scale, $ln(r)$, towards an exponential behavior. In this case, 
it is still possible to determine the thermal conductivity using Eq. \ref{eq2} although the error increases substantially  due to the small curvature ($A=P_{abs}/2\pi d a$) 
of the thermal decay. 
Although a clear deviation from a linear behavior is observed as the heating power increases, this effect is negligible at low heating powers and the thermal
decay can be fully explained using Eq. \ref{eq1}.

The main potential of the technique reported here is demonstrated in Figs. \ref{fig2} and \ref{fig3}. The determination of the thermal conductivity reduces simply a fit to a 
straight line in logarithmic scale at low heating powers, which substantially reduces the error. In addition, assuming that the functional dependence of the thermal conductivity on temperature is known, the thermal conductivity can be determined over a wide temperature range. This technique is suitable to investigate a large variety of materials given that they exhibit appreciable Raman signal. In addition to 1- and 2-dimensional materials it is also possible to extend it to the 3-dimensional case provided a spherical symmetry around the heating spot is maintained. Furthermore, this technique can be extended to the case of thin films supported on substrates, although a numerical analysis may be necessary to solve the heat equation.

\section {CONCLUSIONS}

We have developed a novel high resolution contactless technique for thermal conductivity determination and thermal field mapping
based on the concept of Raman thermometry. A two-laser approach is used to create and probe the spatial decay of the thermal
field from which the thermal conductivity is extracted. This technique is suitable to investigate the temperature distribution in 1- and 2-dimensional
structures, and it can be extended to 3-dimensions for surface temperature mapping. The temperature resolution depends on the 
investigated material and it is $\pm 2$ K for Si, whereas its spatial resolution is diffraction limited and can be as low as 300 nm.

The simplest solution for the temperature distribution in the isotropic 2-dimensional case with a temperature independent thermal conductivity
is shown to be $T(r)\propto ln(r)$, and $\partial T(r)/\partial ln(r)=-P_{abs}/2\pi d \kappa_0$. Thus, the thermal conductivity can be determined
by fitting the slope of the spatial decay of the thermal field. More sophisticated models considering $\kappa(T)$ are possible leading to an exponential decay of the
temperature field $T(r) \propto exp[-A\cdot ln(r)]$ and were experimentally demonstrated heating the membranes up to 800 K.

Finally, we describe the conditions that must be fulfilled by the sample under investigation in order to apply this technique. 
This technique should provide an extra step towards a deeper understanding of thermal management in $n$-dimensional materials since
it gives the complete thermal response of a system, $T(r)$, subjected to a thermal perturbation. 

\section{acknowledgements}

The authors acknowledge the financial support from the FP7 projects MERGING (grant
nr. 309150), NANO-RF (grant nr. 318352) and NANOTHERM (grant nr. 318117), the Spanish MICINN projects nanoTHERM (grant nr. CSD2010-0044)
and TAPHOR (MAT2012-31392). E.C.A. gratefully acknowledges a Becas Chile 2010 CONICYT fellowship from the Chilean government.

\bibliographystyle{plain}

\end{document}